\documentclass[aps, 
twocolumn,showpacs,
superscriptaddress,
nofootinbib]{revtex4-1}  
\usepackage{graphicx,multirow}  
\usepackage{dcolumn}   
\usepackage{bm,relsize}       
\usepackage{amssymb, amsmath,amsfonts}
\usepackage{slashed}   
\pdfoutput=1

\usepackage[usenames,dvipsnames,svgnames,table]{xcolor}
\usepackage[linktoc=page,bookmarks=false,colorlinks=false,linkbordercolor=RoyalBlue,citebordercolor=ForestGreen,urlbordercolor=CornflowerBlue]{hyperref}

\def\beq{\begin{equation}}
\def\eeq{\end{equation}}
\def\MDM	{M_{\mathsmaller{\rm DM}}}
\def\DM		{\mathsmaller{\rm DM}}

\def\EW		{\mathsmaller{\rm EW}}
\def\D		{\mathsmaller{\rm D}}
\def\S		{\mathsmaller{\rm S}}
\def\N		{\mathsmaller{\rm N}}
\def\R		{\mathsmaller{\rm R}}
\def\L		{\mathsmaller{\rm L}}
\def\E		{\mathsmaller{\rm E}}
\def\S		{\mathsmaller{\rm S}}
\def\A		{\mathsmaller{\rm A}}
\def\FO		{\mathsmaller{\rm FO}}
\def\VA	        {\mathsmaller{V_{\rm A}}}
\def\se		{\sigma_e}

\begin{document}
\widetext
\leftline{DESY 20-120}

\title{\Large Dark matter models for the 511 keV galactic line\\
predict keV electron recoils on Earth}
\author{Yohei Ema}
\affiliation{DESY, Notkestra{\ss}e 85, D-22607 Hamburg, Germany}

\author{Filippo Sala}
\affiliation{LPTHE, CNRS\&Sorbonne Universit\'e, 4 Place Jussieu, F-75252 Paris, France}
\author{Ryosuke Sato}
\affiliation{DESY, Notkestra{\ss}e 85, D-22607 Hamburg, Germany}

\begin{abstract}
We propose models of Dark Matter that account for the 511~keV photon emission from the Galactic Centre, compatibly with experimental constraints and theoretical consistency, and where the relic abundance is achieved via $p$-wave annihilations or, in inelastic models, via co-annihilations.
Due to the Dark Matter component that is inevitably upscattered by the Sun, these models generically predict keV electron recoils at detectors on Earth, and could naturally explain the excess recently reported by the XENON1T collaboration.
The very small number of free parameters make these ideas testable by detectors like XENONnT and Panda-X, by accelerators like NA64 and LDMX, and by cosmological surveys like the Simons observatory and CMB-S4.
As a byproduct of our study, we recast NA64 limits on invisibly decaying dark photons to other particles.

\end{abstract}

\maketitle

\paragraph*{\bf Introduction.}

Data that deviate from standard predictions are lifeblood of progress in physics.
The past few decades have seen a plethora of such observational `anomalies', both in cosmic rays and in underground detectors, that could have been explained by some property of particle Dark Matter (DM).
None of them has been so far enough to claim the discovery of a new DM property, because of the possible alternative explanations in terms of new astrophysical sources, of underestimated systematics, etc, often flavored with a healthy dose of skepticism.
An awareness has therefore emerged that the confirmation of a DM origin for some anomaly would require, as a necessary condition, that many anomalies are intimately linked together within a single model of DM.

It is the purpose of this letter to point out one such link. Not only we propose DM models that explain the observed 511~keV line from the Galactic Centre (GC)~\cite{Prantzos:2010wi,Siegert:2015knp,Kierans:2019aqz}, but also we show they predict electron recoils with energies of the order of a keV, of the right intensity and spectrum to be observed by XENON1T~\cite{Aprile:2019xxb,Aprile:2020tmw} and to explain the excess seen in~\cite{Aprile:2020tmw}.
Our spirit in writing this paper is not to abandon the skepticism praised above, but rather to add an interesting --in our opinion-- piece of information to the debates surrounding both datasets.

\paragraph*{\bf The 511~keV galactic line.}
A 511~keV photon line emission in the galaxy has been observed since the 70's, recent measurements include that with the SPI spectrometer on the INTEGRAL observatory~\cite{Siegert:2015knp} and the one with the COSI balloon telescope~\cite{Kierans:2019aqz}, see~\cite{Prantzos:2010wi} for an earlier review.
The signal displays two components of comparable intensity, one along the galactic disk and one in the bulge, the latter with an extension of $O(10^\circ)$ around the galactic center (GC), strongly peaked, corresponding to a flux of $\simeq 10^{-3}$ photons cm$^{-2}$ sec$^{-1}$~\cite{Siegert:2015knp}.
The line is attributed to the annihilation of $e^+ e^-$ into $\gamma\gamma$ via positronium formation, thus it requires sources injecting positrons in the regions where the emission is seen, and with injection energy smaller than about 3~MeV~\cite{Beacom:2005qv}.

The emission from the galactic disk has been tentatively explained with positron injection from the decay of isotopes coming from nucleosynthesis in stars (see e.g.~\cite{Prantzos:2010wi,Bartels:2018eyb}), while the origin of the emission in the bulge is still the object of debate (`one of the most intriguing problems in high energy astrophysics'~\cite{Prantzos:2010wi}).
Recent proposals to explain the positron injection include, for example, low-mass X-ray binaries~\cite{Bartels:2018eyb} and Neutron Star mergers~\cite{Fuller:2018ttb}.

\paragraph*{\bf The 511 line and Dark Matter: preliminaries.}

Given that the origin of the bulge 511 keV line has not yet been clarified, and given that DM exists in our galaxy, it makes sense to entertain the possibility that the latter is responsible for the former.
A DM origin for the positron injection in the bulge has indeed been investigated since~\cite{Boehm:2003bt}.
The morphology of the signal excludes DM decays in favor of annihilations, see e.g.~\cite{Vincent:2012an}.
The 511~keV line emission in the galactic bulge could be accounted for by self-conjugate DM annihilations into an $e^+e^-$ pair with
\beq
\langle\sigma v\rangle_{511}
\simeq 5\cdot 10^{-31} \Big(\frac{\MDM}{3~\text{MeV}}\Big)^{\!2}\, \frac{\text{cm}^3}{\text{sec}}\,,
\label{eq:fit511}
\eeq
where we have used the best fit provided in~\cite{Vincent:2012an} for an NFW DM density profile, as an indicative benchmark. Different profile shapes and the use of new data for the line could change the precise value of $\langle\sigma v\rangle_{511}$, which however is not crucial for the purpose of this paper.

The need for a positron injection energy smaller than 3~MeV~\cite{Beacom:2005qv} implies that, unless one relies on cascade annihilations~\cite{Jia:2017iyc}, $\MDM \lesssim 3$~MeV.
Since so small values of $\MDM$ have been found to be in conflict with cosmological observations, a simple DM-annihilation origin of the 511~keV line has been claimed excluded in~\cite{Wilkinson:2016gsy}.
Recently, however, the refined analysis of~\cite{Escudero:2018mvt,Sabti:2019mhn} found that values of $\MDM$ down to $\sim 1$~MeV can be made consistent with CMB and BBN, by means of a small extra neutrino injection in the early universe, simultaneous with the electron one from the DM annihilations.
We will rely on this new result in building DM models for the 511~keV line.

Eq.~(\ref{eq:fit511}) clarifies that $s$-wave DM annihilation cannot explain the 511 keV line, because so small cross sections imply overclosure of the universe. To be compatible with a thermal generation of the DM abundance, one therefore needs annihilation cross sections in the early universe much larger than today in the GC.
This is realised for example in two simple pictures, where the DM relic abundance is set by:
\begin{itemize}
\item[$\diamond$] $p$-wave annihilations;
\item[$\diamond$] coannihilations with a slightly heavier partner.
\end{itemize}
We will build explicit DM models that realise each of them in the next two paragraphs.

\medskip

\paragraph*{\bf DM for the 511~keV line: $p$-wave.}

Using $\langle\sigma v\rangle_\text{relic}^{(p)}(\MDM=2~\text{MeV}) \simeq 2.2\cdot 10^{-25} v_\text{rel}^2 \text{cm}^3/\text{sec}$~\cite{Saikawa:2020swg}, we find
\beq
\MDM^{(p)} \simeq 2~\text{MeV}\frac{\langle v_\text{rel}^2\rangle^{1/2}_\text{bulge}}{1.1 \times 10^{-3}}\,,
\label{eq:MDMpwave_FO511}
\eeq
where we have normalised $\langle v_\text{rel}^2\rangle^{1/2}_\text{bulge}$ to the value obtained from the velocity dispersion in the bulge $\sigma \simeq 140$~km/s~\cite{Valenti_2018} \footnote{An interesting future direction would be to refine the DM fit of the excess, by taking into account not only the radial dependence of the DM velocity dispersion (see e.g.~\cite{Ascasibar:2005rw,Rasera:2005sa} for old such studies), but also new data and models for the positron injection from astrophysical sources.},
and where we have assumed that the dominant annihilation channel at freeze-out is $e^+ e^-$.
Note that the preferred DM mass would be the same for non-self-conjugate annihilating DM, for which both $\langle\sigma v\rangle_{511}$ and $\langle\sigma v\rangle_\text{relic}$ are larger by a factor of 2.

\begin{figure}[t]
\includegraphics[width=0.48\textwidth]{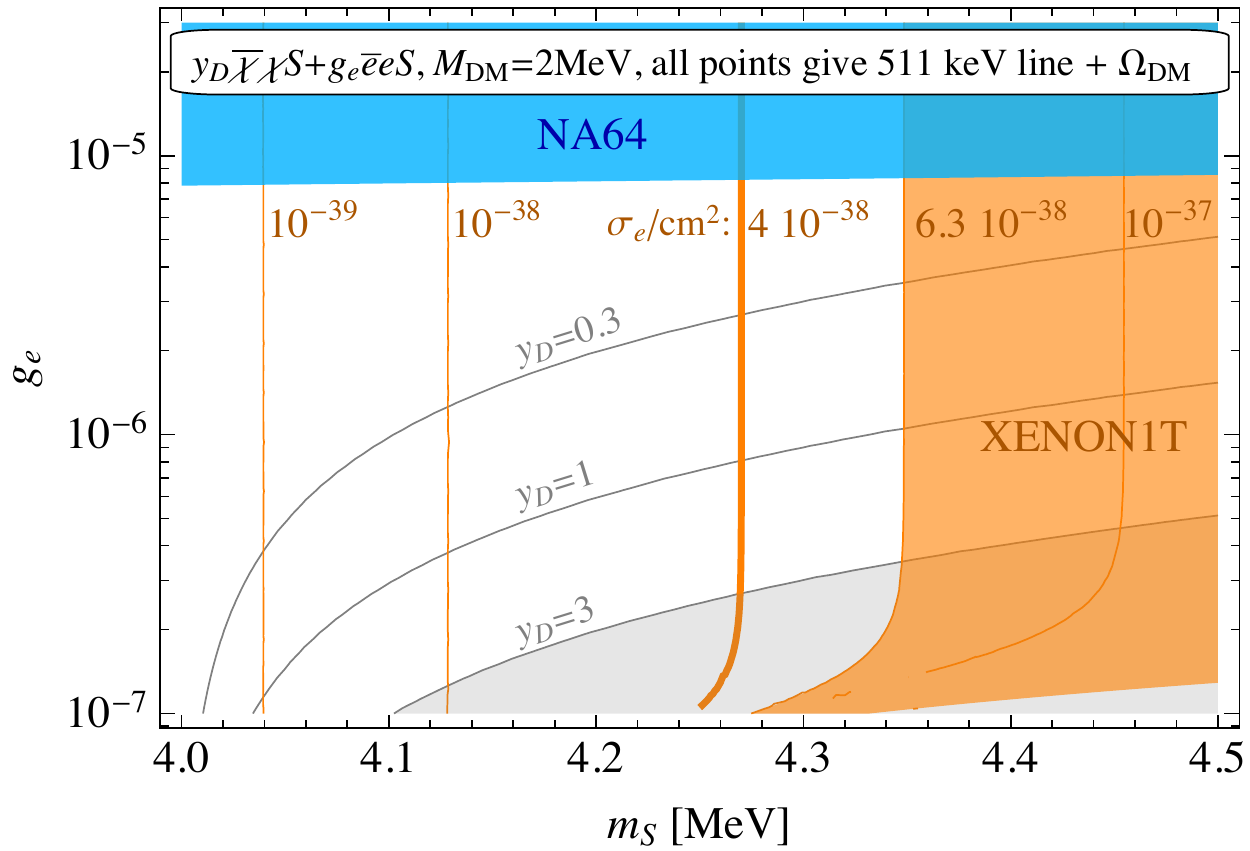}
\caption{\label{fig:pwave_Majorana}
Once the conditions to reproduce the DM abundance and the 511 keV line are imposed, the phenomenology of the model is entirely determined by the scalar mediator mass $m_\S$ and its coupling to electrons $g_e$.
Shaded: non-perturbative dark coupling (gray), our recast of NA64 dark photon limit~\cite{NA64:2019imj} (blue), indicative limit from XENON1T data~\cite{Aprile:2019xxb} (orange). Lines: contours of constant $\sigma_e$ (orange) and of constant dark coupling $y_\D$ (gray).
The thick orange line corresponds to $\se=4\cdot10^{-38}$~cm$^2$, which induces the electron recoil spectrum at XENON1T shown in Fig.~\ref{fig:recoil_spectra}.
}
\end{figure}

An explicit model realising this picture consists of a Majorana fermion $\chi$ as DM candidate, whose interactions with electrons are mediated by a real scalar $S$ via the low-energy Lagrangian (we use 2 component spinor notation throughout this work)
\beq
\mathcal{L}
= y_\D \chi^2 S + g_e e_\L e^\dagger_\R S + \text{h.c.}\,.
\label{eq:L_pwave_simple}
\eeq
This results in the annihilation cross section
\beq
\sigma v_{e^+e^-}
= v_\text{rel}^2
\frac{(y_\D g_e)^2}{8\pi}
\frac{\MDM^2\, \big(1-m_e^2/\MDM^2\big)^{\!\frac{3}{2}}}{\big(m^2_\S-4\MDM^2\big)^2+m^2_\S \, \Gamma^2_\S}\,,
\eeq
and in the cross section for DM-$e$ elastic scattering
\beq
\sigma_e
= \frac{(y_\D g_e)^2}{\pi} 
\frac{\mu_{e\DM}^2}{m^4_\S}\,,
\eeq
where $m_\S$ is the scalar mass, $\Gamma_\S$ its width, and $\mu_{e\DM}=m_e\MDM/(m_e+\MDM)$.
Once $\sigma v_{e^+e^-}$ and $\MDM$ are fixed by the requirements to fit the 511 keV line eq.~(\ref{eq:fit511}) and to reproduce the correct relic abundance eq.~(\ref{eq:MDMpwave_FO511}), then only two free parameters are left, which we choose as $g_e$ and $m_\S$ in Fig.~\ref{fig:pwave_Majorana}.
We find that a region capable of explaining the 511~keV line exists, delimited by perturbativity, direct detection (derived later) and collider limits (see the Appendix~\ref{app:NA64}).\footnote{Limits from CMB~\cite{Slatyer:2015jla}, CR electrons~\cite{Boudaud:2018oya} and CR-electron-upscattered DM~\cite{Ema:2018bih,Cappiello:2019qsw} do not constrain the explanation of the 511~keV line in the models presented in this paper.}

The existence of 3 degrees of freedom with masses $\MDM$ and $m_\S$ of a few MeV is not in conflict with cosmological data,
provided one posits a small injection of neutrinos in the early universe in a proportion $\sim 1:10^4$ to the electron injection, see~\cite{Escudero:2018mvt,Sabti:2019mhn}.
This can for example be achieved with a coupling to neutrinos, $g_\nu \nu^2 S$, of size $g_\nu \sim 10^{-2} g_e$, and where $g_e \sim 10^{-6}$ in the region allowed by the various limits, see Fig.~\ref{fig:pwave_Majorana}. Coupling of neutrinos and electrons of these sizes can be easily obtained in electroweak-invariant completions of the Lagrangian of eq.~(\ref{eq:L_pwave_simple}).
Since they do not present any particular model-building challenge, we defer their presentation to Appendix~\ref{app:UVcompletions}.
The results of~\cite{Escudero:2018mvt,Sabti:2019mhn} indicate that agreement with cosmological data fixes that ratio up to roughly one order of magnitude, so in this sense we do not need a very precise tuning between the electron and neutrino couplings.

Coming to future tests of this model, direct detection experiments like XENONnT and Panda-X will play a leading role in testing the available parameter space of Fig.\ref{fig:pwave_Majorana}. We stress that the shape of the electron recoil spectrum is fixed over the entire parameter space, only its normalisation changes according to the DD cross section shown by the orange lines.
LDMX~\cite{Akesson:2018vlm} will further cut in the available parameter space, as it can probe invisibly decaying dark photon with $m_V = 15~{\rm MeV}$ down to $\epsilon = 10^{-6}$, corresponding to $g_e$ of the same order (see Appendix~\ref{app:UVcompletions}).
Finally, according to Ref.~\cite{Sabti:2019mhn}, both CMB-S4~\cite{Abazajian:2019eic} and the Simons Observatory~\cite{Ade:2018sbj} will probe $\MDM = 2$~MeV at 95\%CL or more and regardless of the ratio of the electron and neutrino couplings, thus offering useful complementary information.

\medskip

\paragraph*{\bf DM for the 511~keV line: coannihilations.}

As a model that concretely realises this idea, we add to the SM a gauge group $U(1)'$, two fermions $\xi$ and $\eta$ with charges 1 and -1 respectively, and a scalar $\phi$ with charge 2 that spontaneously breaks the symmetry.
The most general low-energy Lagrangian that preserves charge conjugation ($\eta \leftrightarrow \xi$, $\phi \leftrightarrow \phi^*$, $V_\mu \leftrightarrow -V_\mu$) reads
\begin{eqnarray}
\mathcal{L} & = & V(|\phi|)
+ \frac{\epsilon}{2} V_{\mu\nu} F^{\mu\nu}
+ (ig_\D \chi_2^\dagger \bar{\sigma}_\mu \chi_1 V^\mu +\text{h.c.}) \nonumber \\
&-& \frac{\bar{m}}{2} (\chi_1^2 + \chi_2^2)
- \frac{y_\phi}{2} (\phi+\phi^*) \big(\chi_2^2 - \chi_1^2 \big)
+\text{h.c.}  \label{eq:L_inelastic}\,,
\end{eqnarray}
where $\chi_1 = i (\eta - \xi)/\sqrt{2}$ and $\chi_2 = (\eta + \xi)/\sqrt{2}$ are the Majorana mass eigenstates, $F_{\mu\nu}$ is the electromagnetic field strength and we have understood all kinetic terms.
The scalar mass and triple-coupling read
\beq
V(|\phi|) = \lambda_\phi \Big( |\phi|^2 - \frac{v_\phi^2}{2} \Big)^{\!2}
\; \Rightarrow \; 
m_\varphi^2 = 2 \lambda_\phi v_\phi^2\,, \;
\lambda_{\varphi^3} = 6 \lambda_\phi v_\phi\,,
\eeq
where $\phi =( \varphi + v_\phi)/\sqrt{2}$ and $\lambda_{\varphi^3}$ is defined by $\mathcal{L}\supset \lambda_{\varphi^3} \varphi^3/6$.
The physical vector and fermion masses read
\beq
m_V = 2 g_\D v_\phi, \qquad
m_{1,2} = \bar{m} \pm \frac{\delta}{2}, \qquad
\delta =2\, \sqrt{2} \, y_\phi v_\phi.
\eeq

$\chi_1$ coannihilates with $\chi_2$ via dark photon exchange.
In the limit $\delta \ll m_{1,2} = \MDM$, one finds
\beq
\sigma v_{\chi_1\chi_2 \to e^+e^-}
=  4 \alpha_e \epsilon^2 g^2_\D \frac{\MDM^2+m_e^2/2}{(m_V^2-4 \MDM^2)^2}\sqrt{1-\frac{m_e^2}{\MDM^2}}\,,
\label{eq:coannihilation}
\eeq
where $\alpha_e$ is the fine-structure constant.
For definiteness, we then assume that $\chi_2$ decays on cosmological scales, such that coannihilations cannot be responsible for a positron injection in the GC today. We will come back to this point in the end of the paragraph.

One can then explain the 511~keV line, if $m_\varphi < \MDM$ and $\varphi$ decays to $e\bar{e}$, via pair annihilations $\chi_i\chi_i \to \varphi\varphi$.
The associated cross section, at first order in $y_\phi v_\phi/\lambda_{\varphi^3} \ll 1$, reads ($i=1,2$) 
\beq
\sigma v_{\chi_i\chi_i \to \varphi\varphi}
= v_\text{rel}^2
\frac{y_\phi^2\,\lambda_{\varphi^3}^2}{64\pi} \frac{1}{(4\,m_i^2 - m_\varphi^2)^2}\sqrt{1-\frac{m_\phi^2}{m_i^2}}\,.
\label{eq:self_annihilation}
\eeq
An operator $|\phi|^2 (e_\L e^\dagger_\R +\text{h.c.})/\Lambda_{\phi e}$
with $\Lambda_{\phi e} \sim 10^{9-10} v_\phi$ guarantees that $\varphi$ decays to $e\bar{e}$ instantaneously on astrophysical scales, while being allowed by collider, supernovae and BBN limits~\cite{Krnjaic:2015mbs,Dev:2020eam}.
It could originate --at the price of some tuning-- from a $|\phi|^2 |H|^2$ term, or from the models discussed in Appendix~\ref{app:UVcompletions}.
Since a $\chi_1\chi_1$ annihilation injects two $e\bar{e}$ pairs, the cross section that best fits the 511~keV line is reduced by a factor of 2 with respect to eq.~(\ref{eq:fit511}).
Therefore we impose
\beq
\sigma v_{\chi_i\chi_i \to \varphi\varphi} = \frac{1}{2} \langle\sigma v\rangle_{511}\,\frac{v_\text{rel}^2}{\langle v_\text{rel}^2\rangle_\text{bulge}}\,.
\label{eq:impose_511_inelastic}
\eeq

If $\chi_i\chi_i \to \varphi\varphi$ were the only processes responsible for the DM abundance, then we would have found another realisation of the $p$-wave annihilating idea, just with $\MDM \simeq 4$~MeV.\footnote{
This is larger than 2~MeV of the previous section because of the factor of 2 with respect to eq.~(\ref{eq:fit511}) that we just explained, and because the relic cross-section is twice that of self-conjugate particles, because $\chi_1\chi_2$ cannot annihilate via $\sigma v_{\chi_i\chi_i \to \varphi\varphi}$.
Note that, for $\MDM < 6$~MeV, the positron injection energy is always smaller than the needed 3~MeV thanks to the extra step in the annihilation.
}
It follows that, for $\MDM \lesssim 4$~MeV, the DM relic density is set dominantly by coannihilations.
We then fix $\epsilon$ by the simple requirement
\beq
\sigma v_{e^+e^-} + 3\frac{\sigma v_{\chi_i\chi_i \to \varphi\varphi}/v_\text{rel}^2}{x_\FO}
= \sigma v^{(s)}_\FO\,,
\label{eq:inelastic_FO}
\eeq
where the left-hand side sums the $s$- and $p$-wave contributions (see e.g.~\cite{Kolb:1990vq} for the origin of the relative factors) and where we use for simplicity the $s$-wave values at $\MDM = 3$~MeV, $\sigma v^{(s)}_\FO \simeq 8\times 10^{-26} \text{cm}^3/\text{sec}$~\cite{Saikawa:2020swg} and $x_\FO \simeq 15$ (their dependence on $\MDM$ is very mild).

\begin{figure}[t]
\includegraphics[width=0.49\textwidth]{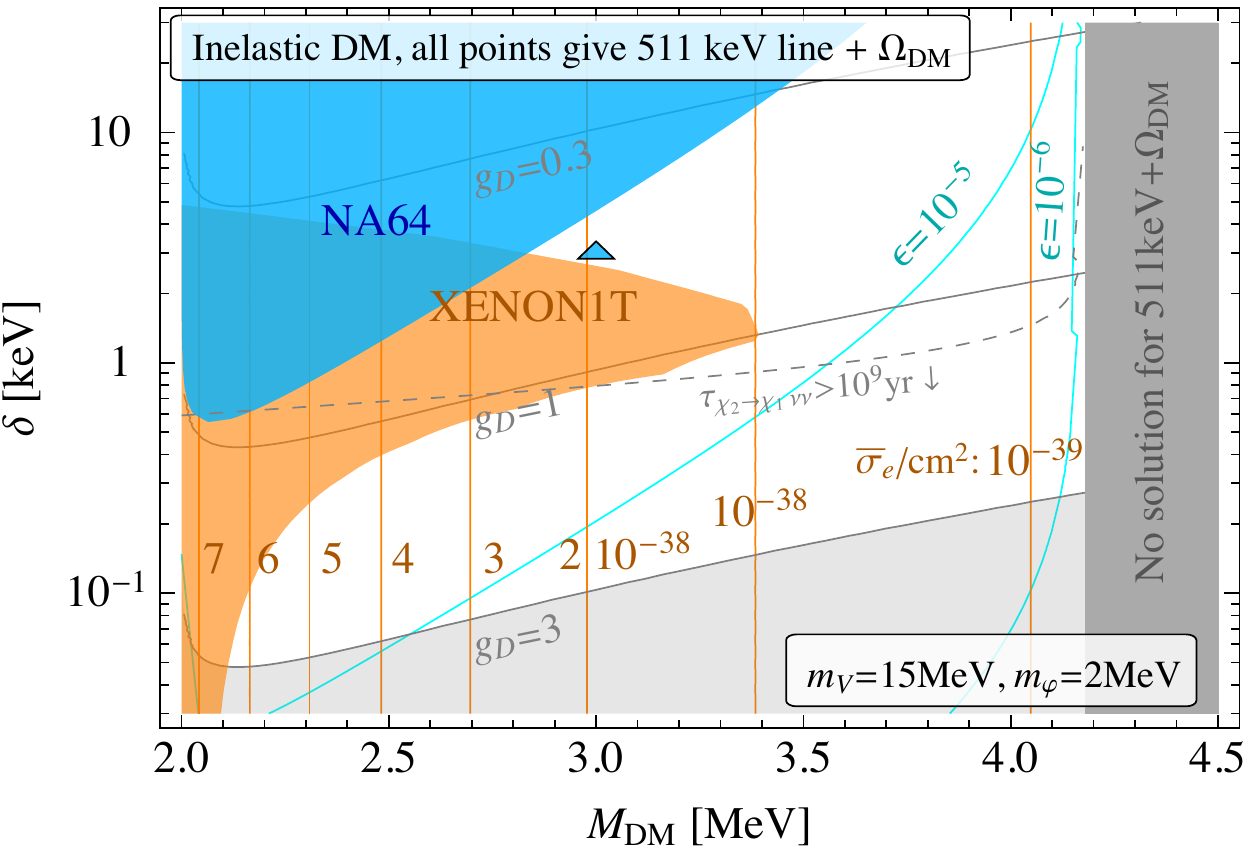}
\caption{\label{fig:coannihilations_chi2decays}
The conditions to reproduce the DM abundance and the 511 keV line impose $\MDM \lesssim 4$~MeV and leave 4 free parameters, chosen here as $\MDM$, $\delta$, $m_\varphi$ and $m_V$.
Shaded: non-perturbative dark coupling (gray), NA64 limit~\cite{NA64:2019imj} (blue), indicative limit from XENON1T data~\cite{Aprile:2020tmw} (orange). Lines: $\bar{\sigma}_e$ (orange), $g_\D$ (gray), $\epsilon$ (cyan).
The dashed gray line roughly delimits the region where $\chi_2$ decays into neutrinos are not enough to deplete the primordial $\chi_2$ population, and further constraints could arise.
The blue triangle corresponds to the electron recoil spectrum at XENON1T shown in Fig.~\ref{fig:recoil_spectra}, and it explains the excess events presented in~\cite{Aprile:2020tmw}.
}
\end{figure}

The model is then left with 4 free parameters, we visualise its parameter space in Fig.~\ref{fig:coannihilations_chi2decays} for the benchmark values $m_\varphi = 2$~MeV and $m_V = 15$~MeV.\footnote{The phenomenology we discuss next is not affected by their precise values, as long as $1.5 \lesssim m_\varphi/\text{MeV} \lesssim~3$, and $10 \lesssim m_V/\text{MeV} \lesssim 100$, where the lower limits are potentially in conflict with BBN and the upper ones close the available parameter space.
Since $\bar{\sigma}_e$ is independent of $m_\varphi$, $m_\varphi < 2$~MeV would not open any new allowed parameter space.}
The allowed region is again delimited by perturbativity, direct detection and collider limits.
Analogously to the previous model, these low values of $\MDM$ can be brought in agreement with BBN and CMB data by a coupling $g_\nu V_\mu \nu^\dagger\bar{\sigma}^\mu\nu$, with $g_\nu \sim 10^{-2} e \epsilon$. We refer the reader to the Appendix~\ref{app:UVcompletions} for a possible origin of $g_\nu$.
Here we just point out that it induces $\Gamma_{\chi_2 \to \chi_1\bar{\nu}\nu} \simeq g^2_\nu g^2_\D\delta^5/(40 \pi^3 m_V^4)$, which for $m_V = 15$~MeV and $\delta \gtrsim 1$~keV implies $\tau_2 < 10^9$~years, so that all $\chi_2$'s left after freeze-out have decayed by today.
Larger values of $\tau_2$ can be avoided by adding another operator to mediate $\chi_2$ decays (e.g. a dipole), otherwise values of $\delta \lesssim 1$~keV could potentially be in conflict with searches for the primordial population of $\chi_2$~\cite{Baryakhtar:2020rwy}.

The allowed values of $\delta$ are restricted around a few keV, which is particularly interesting because they could explain~\cite{Baryakhtar:2020rwy} the excess events at XENON1T~\cite{Aprile:2020tmw}, as we explicitly derive in the next paragraph.
The event rate at XENON1T is proportional to the cross section $\chi_2 e \to \chi_1 e$ in the limit $\delta \to 0$,
\beq
\bar{\sigma}_e = 4 \alpha_e g^2_\D \epsilon^2 \frac{\mu_{e\DM}^2}{m_V^4}\,,
\eeq
which we also display in Fig.~\ref{fig:coannihilations_chi2decays}.

Fig.~\ref{fig:coannihilations_chi2decays} also reports the aforementioned collider limits, and clarifies the impact that future experiments could have in testing this model. The LDMX sensitivity to dark photons will allow to almost completely probe the available parameter space. Cosmological surveys and especially DD experiments will be sensitive to a sizeable chunk of the parameter space, and thus will play an important complementary role in confirming or refuting our interpretation of the 511~keV GC line.

Finally, we left out of this study the case where there is a residual population of $\chi_2$ today, which has also been shown to possibly explain the excess events at XENON1T~\cite{Harigaya:2020ckz,Lee:2020wmh,Bramante:2020zos,Baryakhtar:2020rwy,Bloch:2020uzh,An:2020tcg,Baek:2020owl,He:2020wjs}.
While this goes beyond the purpose of this work, it would be interesting to investigate it in combination with the 511~keV line and we plan to come back to it in future work.

\medskip

\paragraph*{\bf keV electron recoils from Sun-upscattered DM.}
\label{sec:recoils}
The models we proposed to explain the 511~keV line require DM with a mass of a few MeV, interacting with electrons.
Such a DM is efficiently heated inside the sun, resulting in a flux of solar-reflected DM with kinetic energy ($\sim \mathrm{keV}$) significantly larger than the one of halo DM, thus offering new detection avenues to direct detection experiments~\cite{An:2017ojc}.
We now show that, via this higher-energy component, both `$p$-wave' and `coannihilations' models for the 511~keV line automatically induce electron-recoil signals that are probed by XENON1T S2-only~\cite{Aprile:2019xxb}  and S1$+$S2~\cite{Aprile:2020tmw} data.

We outline the procedure to obtain the event rate caused by the solar-reflected DM flux and refer to the Appendix~\ref{app:solar} for more details.
In the case of our interest with relatively small $\sigma_e$, the solar-reflected DM flux $\Phi_\mathrm{refl}$ is estimated as
\begin{align}
	\frac{d \Phi_\mathrm{refl}}{dE} \simeq 
	\frac{n_\DM}{\left(1 \mathrm{AU}\right)^2}
	\int_0^{r_\mathrm{sun}} \!\! dr\,r^2
	\frac{v_\mathrm{esc}(r)}{v_\DM}\,
	n_e(r)
	\left\langle \frac{d\sigma_e}{dE} v_e (r)\right\rangle,
	\label{eq:fluxDM}
\end{align}
where $E$ is the DM kinetic energy, $n_\DM$ is the DM 
number density,
$r_\mathrm{sun}$ is the solar radius, $v_\mathrm{esc}$ is the escape velocity,
$v_\DM$ is the halo DM velocity,
$n_e$($v_e$) is the electron number density (velocity),
and $\langle ... \rangle$ denotes the thermal average.
In this formula, we have improved the analysis of~\cite{Baryakhtar:2020rwy} by including the radial dependence of the solar parameters, taken from~\cite{Bahcall:2004pz}.
The recoil spectrum of the electron initially in the $(n, l)$ state of a XENON atom is given by
\begin{align}
	\frac{dR_{nl}}{dE_R} &= 
	\frac{N_T {\sigma}_e}{8\mu_{e\DM}^2 E_R}\int dq\,q \left\lvert f_{nl}\right\rvert^2
	\xi\left(E_\mathrm{min}\right), 
	\label{eq:devent1} \\
	\xi\left(E_\mathrm{min}\right) &= \int_{E_\mathrm{min}} \!\!dE\,
	\left(\frac{\MDM}{2E}\right)
	\frac{d\Phi_\mathrm{refl}}{dE}, 
	\label{eq:devent2} \\
	E_\mathrm{min} &= \frac{\MDM}{2}\left(\frac{E_{nl} + E_R - \delta}{q} 
	+ \frac{q}{2 \MDM}\right)^2,
	\label{eq:devent3}
\end{align}
where $N_T$ is the number of target particles and $E_{nl}$ is the electron binding energy, see e.g.~\cite{Essig:2015cda} for a detailed derivation of the above expressions.
We compute the atomic form factor $f_{nl}$
following~\cite{Essig:2011nj,Bloch:2020uzh},
and leave a refined treatment including relativistic effects~\cite{Roberts:2016xfw,Roberts:2019chv} to future work.

\begin{figure}[t]
\includegraphics[width=0.49\textwidth]{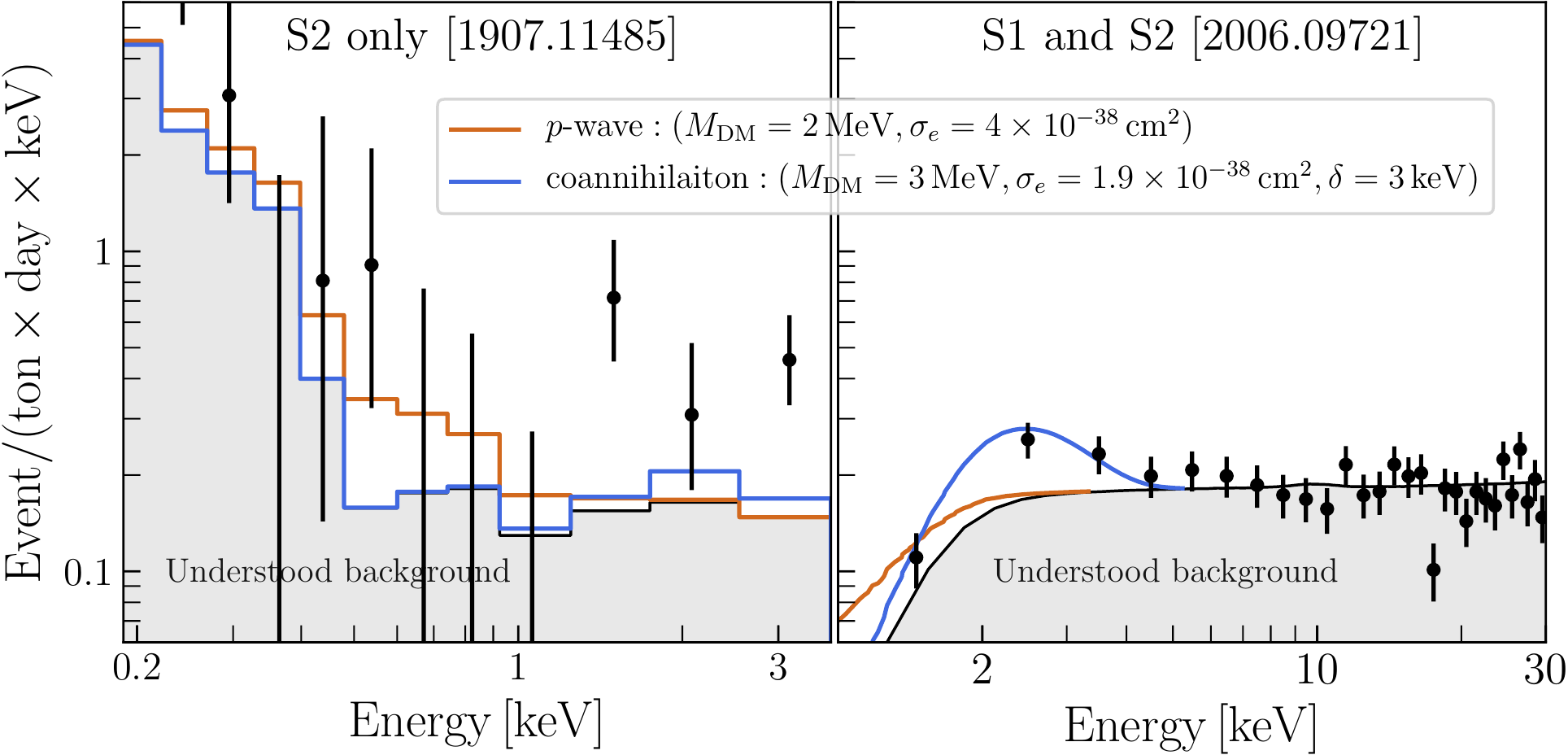}
\caption{\label{fig:recoil_spectra}
Electron recoil spectra induced by solar-upscattered DM, for two benchmark values of the parameters of models that explain the 511~keV line. We overlay them with data and expected backgrounds from the XENON1T S2~\cite{Aprile:2019xxb} (left) and S1+S2~\cite{Aprile:2020tmw} (right) analyses.
}
\end{figure}

In Fig.~\ref{fig:recoil_spectra}, we show the electron recoil spectra for two benchmark points
$\MDM = 2\,\mathrm{MeV}$ and $\sigma_e = 4\times 10^{-38}\,\mathrm{cm}^2$ in the $p$-wave case and
$\MDM = 3\,\mathrm{MeV}$, $\sigma_e = 1.9\times 10^{-38}\,\mathrm{cm}^2$ and $\delta = 3\,\mathrm{keV}$ in the coannihilation case.
The induced electron recoils peak at energies below 2~keV in the $p$-wave case, and in the coannihilation one if $\delta \lesssim$~keV.
In the former case, the position of the peak is fixed by the dark matter mass Eq.~(\ref{eq:MDMpwave_FO511}),
and it does not appear possible to explain signal excess observed at XENON1T. On the other hand,
in the latter case with larger $\delta$ the events instead peak at $E_R \sim \delta$, because the downscattering $\chi_2 \rightarrow \chi_1$ releases more energy than the initial one of $\chi_2$.
In particular, the events are peaked at
$E_R = 2$--$3\,\mathrm{keV}$ in our benchmark point, 
which can explain the recent XENON1T anomaly.
We emphasize that this result is non-trivial, because the allowed parameter region is defined by requirements and experimental limits that are completely independent of XENON1T.
It is then a fortunate accident that this region is in the right ballpark for the explanation of the XENON1T anomaly.

The results of this paragraph are of course interesting beyond these anomalies, as they quantify how XENON1T tests models of light electrophilic DM.
The limits shown in Figs.~\ref{fig:pwave_Majorana} and~\ref{fig:coannihilations_chi2decays} are derived by the conservative requirement that signal plus background should not overshoot the data in~\cite{Aprile:2020tmw} by more than 3$\sigma$, a more precise limit derivation is left to future work.

\medskip

\paragraph*{\bf Conclusions and Outlook.}

We have presented two models which explain the 511 keV line in the galactic bulge by annihilation of particle dark matter with a mass of order MeV. The relic abundance is set by p-wave annihilations in one model, and by coannihilations with a slightly heavier partner in the other model. We have found the novel result that these models induce electron recoils on Earth that are being tested by XENON1T, and that coannihilationmodels could, non-trivially, simultaneously explain the 511 keV line and the excess events recently presented by XENON1T \cite{Aprile:2020tmw}.  In addition, we have demonstrated that both models are compatible with all experimental constraints, in particular with cosmological ones: to evade the conclusion of~\cite{Wilkinson:2016gsy} that no ${\cal O}({\rm MeV})$ DM model could explain the 511~keV line, we have relied on an extra annihilation channel into neutrinos and on the new results of~\cite{Escudero:2018mvt,Sabti:2019mhn}.

Independently of the XENON1T anomaly, our proposed DM explanations of the 511~keV constitute a new physics case for experiments sensitive to keV electron recoils, like XENONnT and Panda-X~\cite{Fu:2017lfc}, for accelerators like NA64 and LDMX~\cite{Akesson:2018vlm}, and for cosmological surveys like CMB-S4~\cite{Abazajian:2019eic} and the Simons Observatory~\cite{Ade:2018sbj}.
The origin of a long-standing astrophysical mystery could be awaiting discovery in their data.

\medskip

\subsection*{Acknowledgements}

We thank Marco Cirelli, Simon Knapen, Yuichiro Nakai, Diego Redigolo and Joe Silk for useful discussions.
\medskip

{\footnotesize
\noindent Funding and research infrastructure acknowledgements: 
\begin{itemize}
\item[$\ast$] Y.E. and R.S. are partially supported by the Deutsche Forschungsgemeinschaft under Germany's Excellence Strategy -- EXC 2121 ``Quantum Universe'' - 390833306;
\item[$\ast$] F.S. is supported in part by a grant ``Tremplin nouveaux entrants et nouvelles entrantes de la FSI''.
\end{itemize}
}

\appendix

\section{Recast of NA64 limits.}
\label{app:NA64}
\begin{figure}[t]
\includegraphics[width=0.49\textwidth]{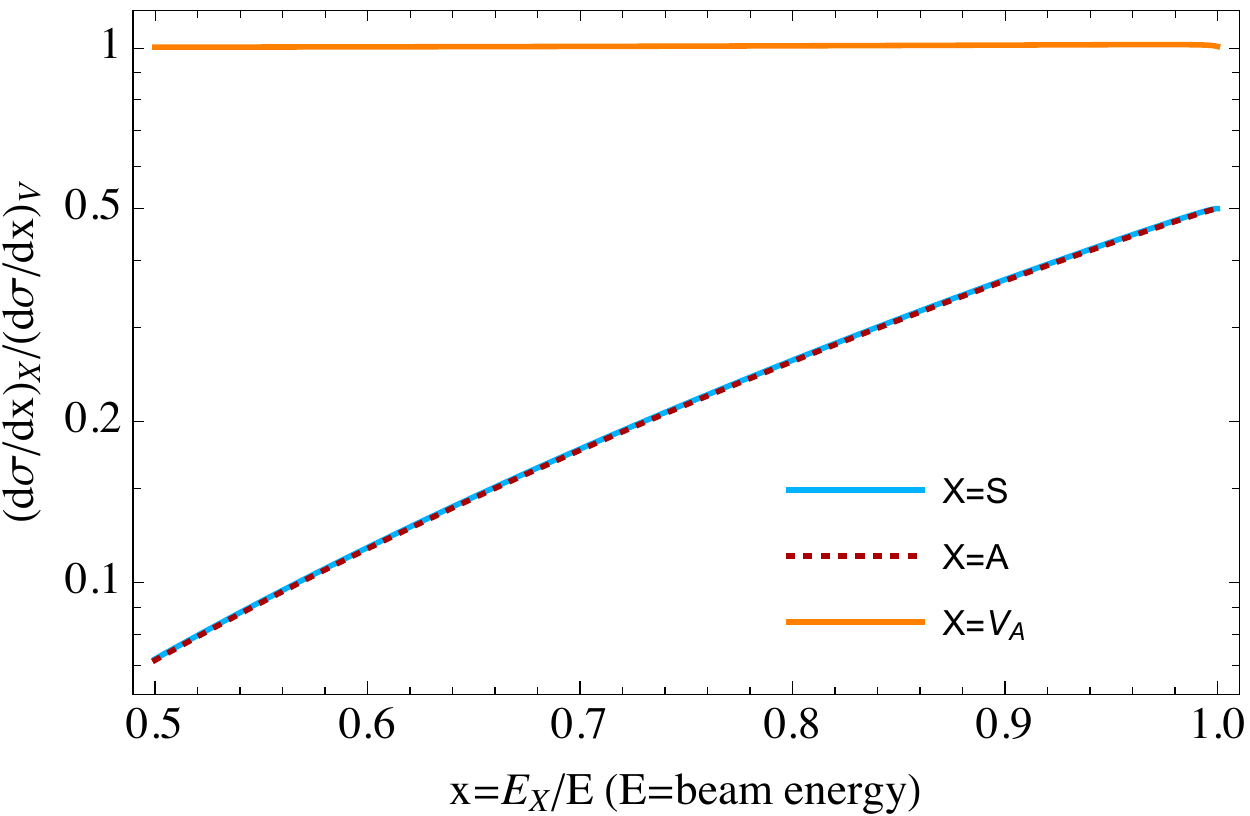}
\caption{\label{fig:IWWcrosssections}
Ratios of cross sections $d\sigma(eZ\to eZX)/dx$, with $x = E_X/E_\text{beam}$ ($E_\text{beam} =100$~GeV for NA64).
Numerator: $X=S$ (blue), $A$ (dotted-red), $V_A$ (orange); denominator: $X = V$.
The range $x \geq 0.5$ is the one relevant for the NA64 searches~\cite{NA64:2019imj} that we are recasting here.
We use the cross sections in the ``improved Weizsaecker-Williams'' approximations as given in~\cite{Liu:2016mqv,Liu:2017htz}.
All curves assume $m_X = 10$~MeV, the dependence on $m_X$ is within the thickness of each line for $m_X > 8$~MeV, and within $\sim 20\%$ of each line for $m_X > 3$~MeV.
}
\end{figure}

NA64 sets the strongest existing constraints on invisibly decaying dark photons in~\cite{NA64:2019imj}: the kinetic mixing $\epsilon$, defined as in eq.~\eqref{eq:L_inelastic}, should be smaller than an $m_V$-dependent function that we denote $\epsilon_\text{limit} (m_V)$.
As we are not aware of any recast of those limits to other invisibly decaying light particles, we perform that recast ourselves, for completeness for  scalars $S$, pseudoscalars $A$ and axial vectors $V_A$, with couplings
\begin{eqnarray}
\mathcal{L}_\S &=& g_e S \, e^\dagger_\L e_\R+\text{h.c.},\\
\mathcal{L}_\A &=& i g_e A \, e^\dagger_\L e_\R+\text{h.c.},\\
\mathcal{L}_\VA &=& i g_e V^\mu_\A \, (e^\dagger_\R \bar{\sigma}_\mu e_\R - e^\dagger_\L \bar{\sigma}_\mu e_\L),
\end{eqnarray}
which in 4-component spinor notation read, respectively, $g_e \bar{e} e S$, $i g_e \bar{e} e A$ and $i g_e \bar{e} \gamma_\mu \gamma_5 e V^\mu_\A$.
We recast NA64 limits by imposing
\beq
g_e (m_{\S,\A,\VA}) < C_{\S,\A,\VA} e \,\epsilon_\text{limit} (m_{\S,\A}),
\label{eq:recast_NA64}
\eeq
where $e$ is the electric charge and
\beq
C_X  = \left(\frac{N_V/(\epsilon^2 e^2)}{N_X /g_e^2}\right)^{\!\frac{1}{2}}\,.
\label{eq:CX}
\eeq
We have defined
\beq
N_X = \int_{0.5}^{x_\text{max}}\!\!dx\, \text{Eff}(x)\,\frac{d\sigma}{dx}(eZ\to eZX)\,,
\label{eq:NX}
\eeq
where $x = E_X/E_\text{beam}$ ($E_\text{beam} =100$~GeV for NA64)
and the lower limit of integration in $x$ comes from the cut $E_\text{miss} > 50$~GeV~\cite{NA64:2019imj}.
The upper limit of integration $x_\text{max}$ satisfies $x_\text{max} < 0.997$, because of the trigger $E_\text{cal} > 0.3$~GeV~\cite{Banerjee:2017hhz}.
For the cross sections $d\sigma(eZ\to eZX)/dx$ we use the ``improved Weizsaecker-Williams'' approximations given in eq.~(33) of~\cite{Liu:2016mqv} for $X=S$ and in eq.~(30) of~\cite{Liu:2017htz} for $X = V,A,V_\A$.
In Fig.~\ref{fig:IWWcrosssections} we display the ratio of the $X=S,A,V_\A$ cross sections and the $X=V$ cross section, the latter being the relevant one for the model on which NA64 has cast its limit.

Finally, the efficiency $\text{Eff}(x)$ has a weak dependence on $x$~\cite{NA64:2019imj}, it does so mostly for $x$ close to one, see the discussion in~\cite{Banerjee:2017hhz} and e.g. Fig.~11 in that paper.
Since we have not found a detailed study of the efficiency of NA64 in the region $x$ close to 1, we assume it is independent of $x$, so that it simplifies in the ratio $N_V/N_X$ that defines our rescaling eq.~(\ref{eq:CX}).
As visible in Fig.~\ref{fig:IWWcrosssections}, this procedure does not introduce any significant error for the axial vector case.
For the scalar and pseudoscalar cases, since the ratios of their cross sections to the vector one are a monotonically increasing function of $x$, and since the efficiency worsens when $x$ approaches one, the value of $C_{\S,\A}$ that we obtain for $x_\text{max} = 0.997$ represent an aggressive estimate of the NA64 exclusion of such particles.
A conservative one can instead be obtained by choosing a value of $x_\text{max}$ below which the efficiency is roughly a constant in $x$, which we take for definiteness as $x_\text{max} = 0.9$.
Our resulting coefficients $C_{\S,\A,\VA}$, for these two extreme limits of integration and for various values of $m_X$, are given in Table~\ref{tab:CX}.
In the $p$-wave model studied in the main text, in order to be conservative on the allowed parameter space, we have used the aggressive rescaling of the NA64 limits, i.e. $C_\S = 1.6$.

\begin{table}
\begin{tabular}{c|cc|cc|cc|l}
$m_X$ [MeV] &  \multicolumn{2}{c}{$C_\S$} &  \multicolumn{2}{c}{$C_\A$} &  \multicolumn{2}{c}{$C_\VA$} &\\
\hline
&  0.997 & 0.9 & 0.997& 0.9 & 0.997 & 0.9 & $x_\text{max}$\\
\hline
 1 & 1.7 & 1.8 & 1.8 & 2.0 & 0.8 & 0.8 &\\
 2 & 1.7 & 2.0 & 1.7 & 2.0 & 0.9 & 0.9 &\\
 3 & 1.6 & 2.0 & 1.7 & 2.1 & 1.0 & 1.0 &\\
 4 & 1.6 & 2.0 & 1.7 & 2.1 & 1.0 & 1.0 &\\
 5 & 1.6 & 2.0 & 1.6 & 2.1 & 1.0 & 1.0 &\\
$\geq 6$ & 1.6 & 2.1 & 1.6 & 2.1 & 1.0 & 1.0 &\\
 \end{tabular}
\caption{\label{tab:CX} Coefficients entering eq.~(\ref{eq:recast_NA64}) to recast NA64 limits, on invisibly decaying dark photons~\cite{NA64:2019imj}, to invisibly decaying scalars S, pseudoscalars A and axial vectors $V_\A$ coupled with electrons as in eq.~(\ref{eq:recast_NA64}). We display our results for two cases of the upper limit of integration $x_\text{max}$ in eq.~(\ref{eq:NX}).}
\end{table}


Another source of uncertainty of our rescaling comes from the fact we used cross sections in the ``improved Weizsaecker-Williams'' approximation.
The comparisons of these cross sections with the full results, in ref.~\cite{Liu:2016mqv,Liu:2017htz}, show that the impact of the approximation over the full $x$ range is analogous for the four cases $X = S,A,V,V_\A$, as one could roughly expect by observing that this approximation consists in a different treatment of the phase-space edges.
Therefore the error in our rescaling, induced by the approximations in the cross section, is qualitatively expected to be smaller than the error in the cross sections themselves, because it relies on ratios.
Since this recast is not the main purpose of this paper, we content ourselves with this procedure, and we encourage the NA64 collaboration to present their very interesting results for particles other than dark photons.

\section{UV completions.}
\label{app:UVcompletions}
We here propose explicit ultraviolet (UV) completions of all the low-energy couplings that are not manifestly electroweak (EW) invariant.

We start by scalar couplings to electrons.
A coupling $g_e$ defined as in eq.~\eqref{eq:L_pwave_simple}, $g_e e_\L e^\dagger_\R S + \text{h.c.}$, of the needed size $g_e \sim 10^{-6}$ (see Fig.~\ref{fig:pwave_Majorana}),
can be obtained by adding to the SM two fermions $E_\L$ and $E^\dagger_\R$, with charge assignments of $e_\R$ and $e^\dagger_\R$ respectively, and Lagrangian
\beq
\mathcal{L}_\E
= y_\E \ell H^\dagger E^\dagger_\R
+M_\E E_\L E^\dagger_\R 
+ g_\E S E_\L e^\dagger_\R\,.
\eeq
This induces a coupling to electrons ($v_\EW \simeq 246$~GeV)
\beq
g_e
\simeq g_\E \frac{y_\E v_\EW}{\sqrt{2} M_\E}
\approx 2 \cdot 10^{-6} y_\E g_\E \frac{10^5~\text{TeV}}{M_\E}\,,
\eeq
which is of the desired size for $M_\E$ out of experimental reach and perturbative values of the  couplings $y_\E$ and $g_\E$.

In the coannihilation model the higher dimensional operator $|\phi|^2 (e_\L e^\dagger_\R +\text{h.c.})/\Lambda_{\phi e}$, that induces the coupling of $\varphi$ to electrons, can be obtained by adding to the SM the fermions $E_\L$ and $E^\dagger_\R$, with SM charge assignments of $e_\R$ and $e^\dagger_\R$ respectively, and $L_\L$ and $L^\dagger_\R$, with SM charge assignments of $\ell$ and $\ell^\dagger$ respectively.
Furthermore, we assign to $E_\L$ and $L_\L$ ($E^\dagger_\R$ and $L^\dagger_\R$) charge $+2$ ($-2$) under the $U(1)'$ gauge group.
The Lagrangian
\begin{eqnarray}
\mathcal{L} &=& M_\L L_\L L^\dagger_\R + M_\E E_\L E^\dagger_\R + y_\L L_\L H^\dagger E^\dagger_\R \nonumber\\
&+& g_\E \phi E_\L e^\dagger_\R + g_\L \phi L_\L \ell^\dagger +\text{h.c.}\,,
\end{eqnarray}
then induces
\beq
\frac{\Lambda_{\phi e}}{v_\phi} \simeq  \frac{M_\L M_\E}{y_\L g_\E g_\L v_\EW v_\phi}
\approx 10^9   \frac{g_\D}{y_\L g_\E g_\L} \frac{15~\text{MeV}}{m_V} \frac{M_\L M_\E}{(40~\text{TeV})^2}\,,
\eeq
where we remind the reader that we needed $\Lambda_{\phi e}/v_\phi$ in the ballpark of $10^{9-10}$, in order for $\varphi$ to decay to $e\bar{e}$ instantaneously on astrophysical scales and compatibly with collider, supernovae and BBN limits~\cite{Krnjaic:2015mbs,Dev:2020eam}.
We have just seen how this value can be achieved by adding new vector-like leptons with masses out of collider reach.

Otherwise the coupling to electrons of both $S$ and $\varphi$ can easily be obtained via operators that mix the new scalars with the Higgs, respectively $S |H|^2$ and $|\phi|^2 |H|^2$.
In the latter case, however, one would need to tune the parameters of $V(|\phi|)$ with this quartic coupling, in order to keep $v_\phi \ll v_\EW$.

\medskip

A coupling of $S$ to neutrinos $g_\nu \nu^2 S$, of size $g_\nu \sim 10^{-2} g_e$ as needed to make the model compatible with cosmological data~\cite{Escudero:2018mvt,Sabti:2019mhn}, can be achieved by extending the SM with three singlet fermions $\nu_\R$, $ N_\L$ and $N_\R$.
The EW-invariant Lagrangian
\beq
\mathcal{L}
= y_\nu \ell H \nu^\dagger_\R + m_\N N_\L N^\dagger_\R + g_\N S N_\L \nu^\dagger_\R + \text{h.c.}\,
\eeq
then induces
\beq
g_\nu \simeq
g_\N \frac{y_\nu v_\EW}{\sqrt{2} M_\N}
\approx 2 \cdot 10^{-8} y_\nu g_\N \frac{10^7~\text{TeV}}{M_\N}\,,
\eeq
which is of the desired size $g_\nu \sim 10^{-2} g_e \sim 10^{-8} $
(see Fig.~1 for the interesting values of $g_e$)
for $N$ out of experimental reach.

\medskip

We finally provide an example of an EW-invariant completion for the small coupling to neutrinos of a $U(1)'$ gauge boson.
We add to the model of eq.~\eqref{eq:L_inelastic} one total singlet fermion $\nu^\dagger_\R$ and two left-handed fermions $N_\L$ and $N_\R$, with charges respectively $+2$ and $-2$ under $U(1)'$, and singlets under the SM gauge group.
The Lagrangian
\beq
\mathcal{L}_\nu =  y_\nu \ell H^\dagger \nu^\dagger_\R + m_\N N^\dagger_\L N_\R + y_\N N_\L \nu_R^\dagger \phi + \text{h.c.}\,
\eeq
then induces a coupling  of size
\beq
g_\nu
\simeq 2\,g_\D \Big(\frac{y_\N v_\phi}{m_\N}\Big)^2
\approx 10^{-7} \frac{y^2_\N}{g_\D}
\Big(\frac{m_V}{15~\text{MeV}}\Big)^{\!2}
\Big(\frac{30~\text{GeV}}{m_\N}\Big)^{\!2}\,.
\eeq
One can then obtain the needed value $g_\nu \sim 10^{-2} e \epsilon \sim 10^{-7}$
(see Fig.~\ref{fig:coannihilations_chi2decays} for the interesting values of $\epsilon$)
for $m_\N \sim 30$~GeV, which is out of experimental reach because $N$ is a total SM singlet.

\section{Solar-reflected DM events at XENON1T.}
\label{app:solar}

Here we give the procedure to compute the electron recoil spectra
at XENON1T in detail.

\begin{figure}[t]
\includegraphics[width=0.49\textwidth]{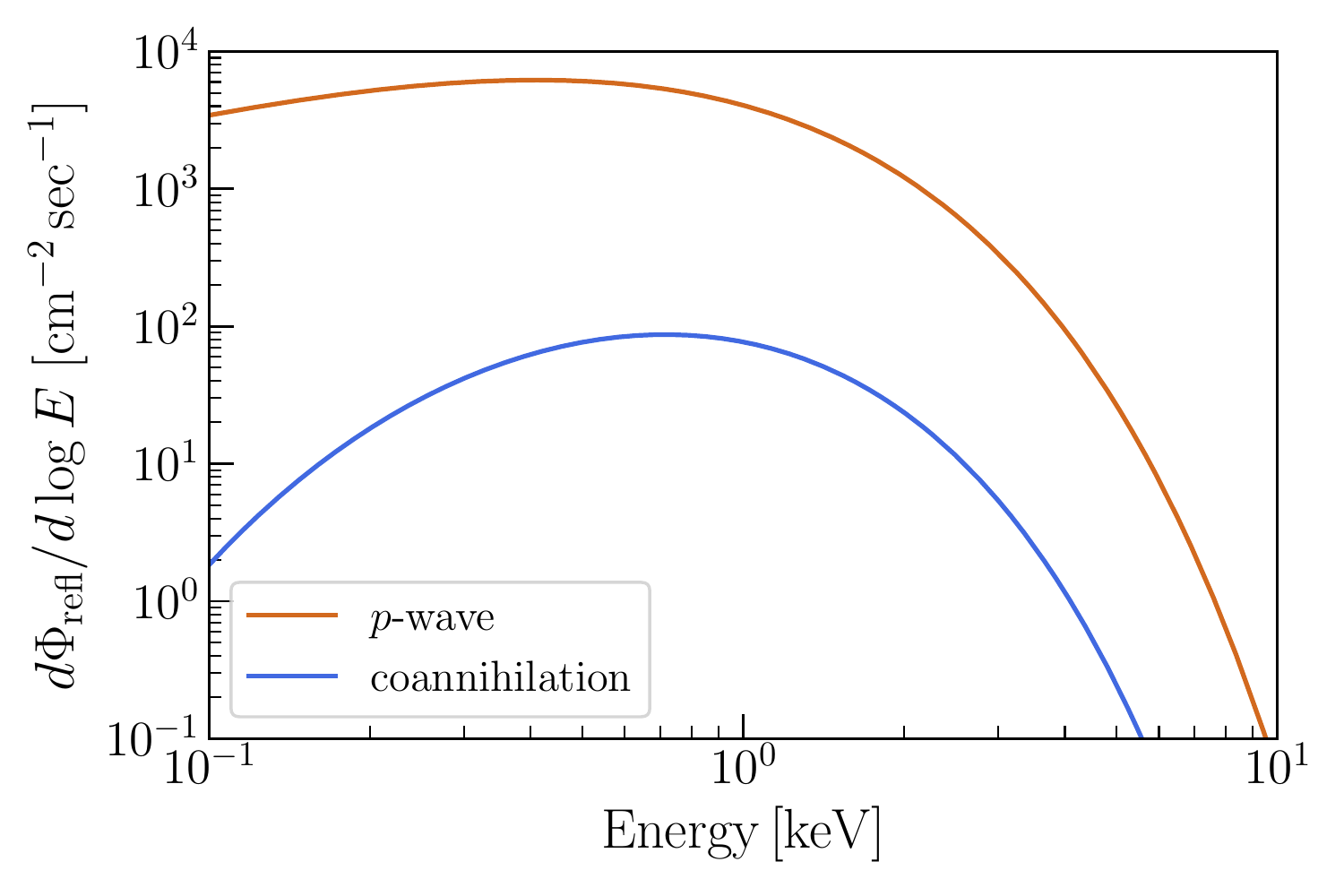}
\caption{\label{fig:fluxDM}
Solar-reflected DM flux
for our benchmark points:
$\MDM = 2\,\mathrm{MeV}$ and $\sigma_e =4\times 10^{-38}\,\mathrm{cm}^2$ in the $p$-wave case and
$\MDM = 3\,\mathrm{MeV}$, $\sigma_e = 1.9\times 10^{-38}\,\mathrm{cm}^2$ and $\delta = 3\,\mathrm{keV}$ in the coannihilation case.
}
\end{figure}

The solar-reflected DM flux is given by eq.~\eqref{eq:fluxDM},
and we explain each term in the following.
We take the DM number density as 
$n_\mathrm{DM} = (0.42\,\mathrm{GeV}/\MDM)\,\mathrm{cm}^{-3}$~\cite{Pato:2015dua,Buch:2018qdr}.
The astrophysical unit is given by $1\,\mathrm{AU} \simeq 1.5\times 10^{13}\,\mathrm{cm}$.
The escape velocity is given by
\begin{align}
	v_\mathrm{esc}(r) = \sqrt{\frac{2GM(r)}{r}},
\end{align}
where $G$ is the Newton constant and $M(r)$ is the solar mass inside the radius $r$.
The factor $v_\mathrm{esc}/v_\mathrm{DM}$ originates from the combination of the enhanced classical cross section by the attractive gravitational potential and the spreading of the flux by the increased DM velocity~\cite{Baryakhtar:2020rwy}.\footnote{
	Precisely speaking, the enhancement of the cross section by the factor 
	$v_\mathrm{esc}^2/v_\mathrm{DM}^2$ applies only when the potential is proportional to $1/r$.
	It is however enough for our purpose, given the uncertainties in the other factors such 
	as the atomic form factor.
}
The halo DM velocity is taken as $v_\mathrm{DM} = 220\,\mathrm{km}/\mathrm{sec}$.
Assuming the Maxwell-Boltzmann distribution,
the thermal averaged differential cross section is given by
\begin{align}
	\left\langle \frac{d\sigma_e}{dE} v_e\right\rangle
	&= \frac{{\sigma}_e \MDM}{\mu_{e\DM}^2} \sqrt{\frac{m_e}{2\pi T}} 
	\exp\left[- \frac{m_e v_\mathrm{min}^2}{2T}\right], \nonumber \\
	v_\mathrm{min} &= \frac{1}{\sqrt{2\MDM E}}\left[\frac{\MDM E}{\mu_{e\DM}} 
	+ \delta\right].
\end{align}
Finally we shift the DM kinetic energy after scattering, by the gravitational potential
at the point of the scattering to take into account the gravitational redshift effect, $E \to E - \MDM v^2_\mathrm{esc}(r)/2$.
In Fig.~\ref{fig:fluxDM}, we show the solar-reflected DM flux
for the benchmark points used in the main text:
$\MDM = 2\,\mathrm{MeV}$ and $\sigma_e = 4\times 10^{-38}\,\mathrm{cm}^2$ in the $p$-wave case and
$\MDM = 3\,\mathrm{MeV}$, $\sigma_e = 1.9\times 10^{-38}\,\mathrm{cm}^2$ and $\delta = 3\,\mathrm{keV}$ in the coannihilation case.

\begin{figure}[t]
\includegraphics[width=0.49\textwidth]{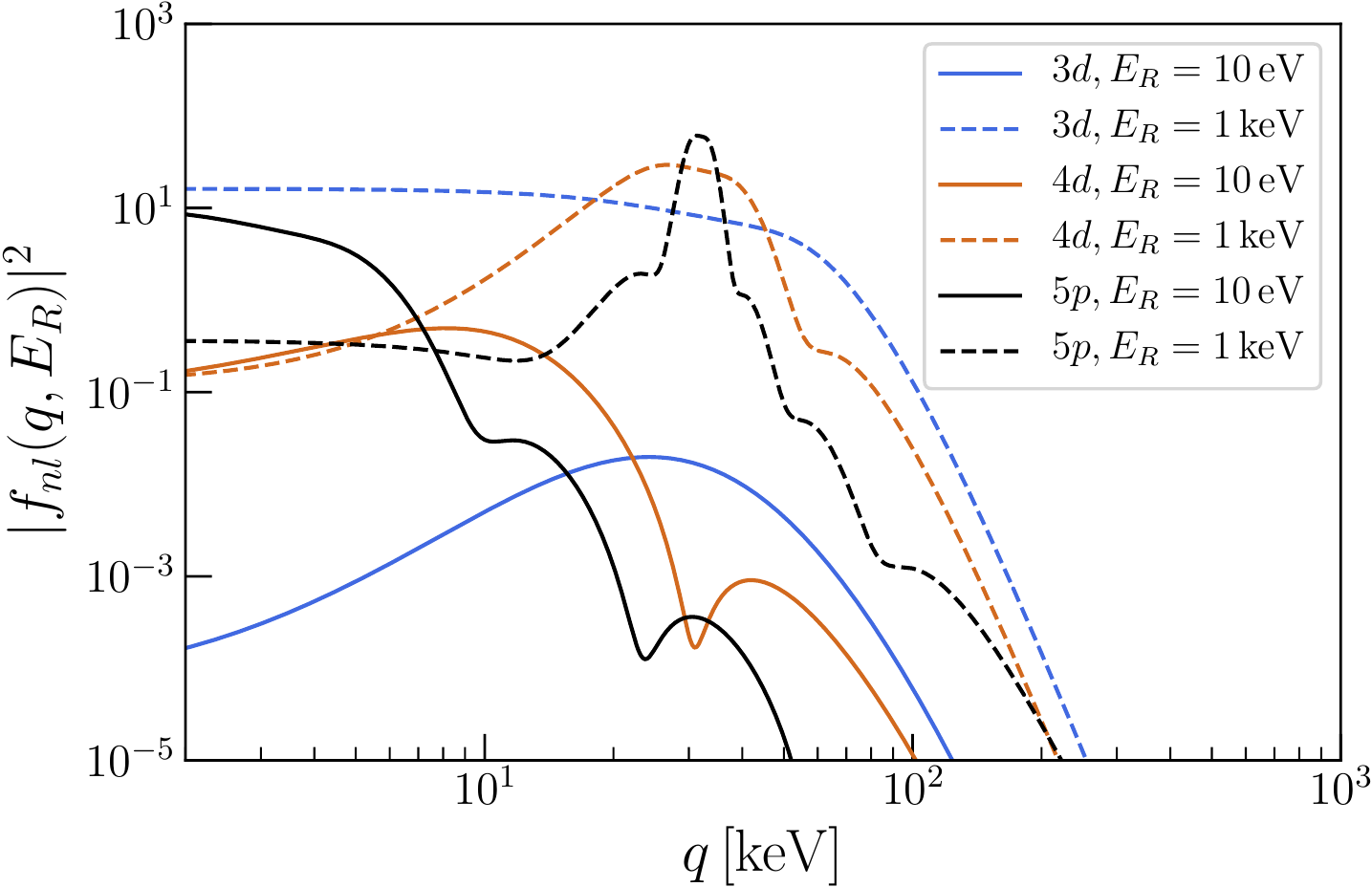}
\caption{\label{fig:atomicFF}
Atomic form factors of the $3d$-, $4d$- and $5p$-state electrons 
without the Fermi factor,
with two different values of the recoil energy $E_R$.
}
\end{figure}

Once the reflected DM flux is computed,
the electron recoil spectra are given by eqs.~\eqref{eq:devent1}--\eqref{eq:devent3},\footnote{
	We think that there is a typo in the formula of $\eta$ in~\cite{An:2017ojc}
	(which is our $\xi$ divided by the total halo DM flux).
} with the number of the target particle taken as $N_T = 4.2\times 10^{27}$ per tonne
in our computation.
As mentioned in the main text, we compute the atomic form factor following~\cite{Essig:2011nj,Bloch:2020uzh}.
Assuming the plane wave function for the out-going electron,
the atomic form factor is given by
\begin{align}
	\left\lvert f_{nl}(q, E_R)\right\rvert^2 
	&= F_\mathrm{Fermi}\frac{2l+1}{2\pi^3}\frac{m_e E_R}{q} 
	\left[\int_{k_-}^{k_+} dk\,k \left\lvert \chi_{nl}\left(k\right)\right\rvert^2\right], \nonumber \\
	k_{\pm} &= \left\lvert \sqrt{2 m_e E_R} \pm q\right\rvert.
\end{align}
The radial part of the wave function in the momentum space $\chi_{nl}$ 
is given by
\begin{align}
	\chi_{nl}(k) = 4\pi \int_0^{\infty} dr\,r^2 j_l\left(kr\right) R_{nl}\left(r\right),
\end{align}
where $j_l$ is the spherical Bessel function and $R_{nl}$ is the radial part of the real space wave function.
We take $R_{nl}$ as
\begin{align}
	R_{nl} &= \sum_{j} C_{jln} N_{jl} r^{n_{jl}-1} \exp\left(-Z_{jl}r\right), \nonumber \\
	N_{jl} &= \frac{\left(2Z_{jl}\right)^{n_{jl} + 1/2}}{\sqrt{\left(2n_{jl}\right)!}}.
\end{align}
where $C_{jln}, Z_{jl}$ and $n_{jl}$ are taken from~\cite{Bunge:1993jsz}.
If we define
\begin{align}
	f_{l}\left(n; x\right) \equiv \frac{2^{n+1/2}}{\sqrt{\left(2n\right)!}}
	\int_0^\infty dy\,y^{n+1} j_{l}\left(xy\right) \exp\left(-y\right),
	\label{eq:integral_basis_function}
\end{align}
the momentum-space wave function is given by
\begin{align}
	\chi_{nl}\left(k \right) = 4\pi \sum_{j} \frac{C_{jln}}{Z_{jl}^{3/2}} f_l\left(n_{jl}; k/Z_{jl}\right).
\end{align}
The integral~\eqref{eq:integral_basis_function} can be analytically performed,
which simplifies the numerical computation.
The wave functions are normalized as
\begin{align}
	\int dk\,k^2 \left\lvert \chi_{nl} \right\rvert^2 = \left(2\pi\right)^3,
	\quad
	\int_0^\infty dr\,r^2 \left\lvert R_{nl}\right\rvert^2 = 1,
\end{align}
which agrees with the normalization of~\cite{Bunge:1993jsz}.
Finally the Fermi factor is given by
\begin{align}
	F_\mathrm{Fermi}(q) = \frac{2\pi d}{1-e^{-2\pi d}},
	\quad
	d = Z_\mathrm{eff} \frac{\alpha_e m_e}{q},
\end{align}
where we take the effective charge as $Z_\mathrm{eff} = 1$.
We show the form factors without the Fermi factor in Fig.~\ref{fig:atomicFF}.
They agree well with~\cite{Bloch:2020uzh} except 
in the region $q \lesssim 10\,\mathrm{keV}$ for the $4d$-state electron
with $E_R = 1\,\mathrm{keV}$, whose effect on the final result is anyway minor.
In our computation we neglect the contribution from $1s$, $2s$ and $2p$ electrons,
because their binding energies are larger than $\simeq 4.8$~keV (see e.g.~\cite{Bunge:1993jsz}) and thus can
be neglected in this specific study. We included 8 orbits, from $3s$ up to $5p$.

After computing the electron recoil spectra,
we convolute them with the detector response to obtain the signals.
For the S2-only analysis, we use the mean values in~\cite{Aprile:2019xxb}
to translate the recoil energy to photoelectron (PE).
Although the efficiency depends on the position of the event,
we simply multiply all the efficiency shown in~\cite{Aprile:2019xxb} to obtain the signals
in this work. A more detailed analysis on the detector response is left as a future work.
For the recent S1$+$S2 analysis, we follow the procedure outlined in the original paper~\cite{Aprile:2020tmw}.
We smear the events by a gaussian distribution with the width given by
\begin{align}
	\sigma\left(E\right) = a \sqrt{E} + b E, 
\end{align}
where we take $a = 0.31 \sqrt{\mathrm{keV}}$ and $b = 0.0037$ in our numerical computation.
We then multiply the efficiency that is again given in~\cite{Aprile:2020tmw}.

\bibliography{511keVline_keVrecoils}

\end{document}